\documentclass[%
 reprint,
 superscriptaddress,
 amsmath,amssymb,
 aps,
]{revtex4-1}

\usepackage{graphicx}
\usepackage{dcolumn}
\usepackage{multirow}
\usepackage{bm}
\usepackage{xcolor}
\usepackage{ulem}
\usepackage{mathrsfs}
\usepackage{tikz}  

\begin{document}


\title{Neutron Double-Differential Cross Sections for Spallation Reactions from an ANN Model}

\author{Rong Wang}
\email{rwang@impcas.ac.cn}
\affiliation{Institute of Modern Physics, Chinese Academy of Sciences, Lanzhou 730000, China}
\affiliation{School of Nuclear Science and Technology, University of Chinese Academy of Sciences, Beijing 100049, China}

\author{Sheng-Ting Sun}
\affiliation{School of Nuclear Science and Technology, Lanzhou University, Lanzhou 730000, China}

\author{Han-Jie Cai}
\email{caihj@impcas.ac.cn}
\affiliation{Institute of Modern Physics, Chinese Academy of Sciences, Lanzhou 730000, China}
\affiliation{School of Nuclear Science and Technology, University of Chinese Academy of Sciences, Beijing 100049, China}

\author{Xun-Chao Zhang}
\affiliation{Institute of Modern Physics, Chinese Academy of Sciences, Lanzhou 730000, China}
\affiliation{School of Nuclear Science and Technology, University of Chinese Academy of Sciences, Beijing 100049, China}

\author{Huan Jia}
\affiliation{Institute of Modern Physics, Chinese Academy of Sciences, Lanzhou 730000, China}
\affiliation{School of Nuclear Science and Technology, University of Chinese Academy of Sciences, Beijing 100049, China}

\author{Yuan He}
\affiliation{Institute of Modern Physics, Chinese Academy of Sciences, Lanzhou 730000, China}
\affiliation{School of Nuclear Science and Technology, University of Chinese Academy of Sciences, Beijing 100049, China}


\date{\today}

\begin{abstract}
In this paper, we present a data-driven artificial neural network (ANN) model for describing the double differential cross sections (DDCS) of neutron emission in nuclear spallation reactions. The ANN model is found to be precise, flexible, and efficient in predicting differential cross sections of nuclear reactions and in learning the complex dependence of neutron DDCS on the projectile energy ($T_p$), target nucleus ($A$ and $Z$), neutron energy ($T_n$), and neutron emission angle ($\theta_n$). The model is trained on replicas of experimental data that incorporate uncertainties. Several regularization schemes are examined during ANN training. The input variables of the constructed ANN framework are also investigated, and the following six key variables are selected for the input layer of the ANN model: $\theta_{n}$, ${\rm log}(T_n/T_p)$, $T_n/T_p$, ${\rm log}(T_p)$, $A^{2/3}$, and $N/Z$. The ANN predictions are compared with training data provided by various experimental collaborations, showing excellent agreement. The resulting model is further tested on test data with projectile energies, target nuclei, and neutron emission angles different from those in the training data, indicating strong predictive power and generalization capability of the ANN framework. As an illustration, the neutron DDCS as functions of $T_n$, $\theta_n$, and projectile energy $T_p$ are predicted and presented for copper target. The proposed high-precision ANN model is expected to be beneficial for accelerator-driven system (ADS) design and many other applications in nuclear physics, astrophysics, and nuclear technology development.
\end{abstract}

\maketitle


\section{Introduction}\label{sec:intro}

The Accelerator-Driven Advanced Nuclear Energy System (ADANES) \cite{Yan2017Concept,He2025Advances} is an innovative ``beyond-Generation-IV'' nuclear energy concept proposed by the Chinese Academy of Sciences, designed to fundamentally address the problems of low uranium utilization ($<1$\%) and long-lived radioactive waste in traditional reactors. Its closed-fuel-cycle scheme, integrating an Accelerator-Driven Subcritical System and a dry fuel reprocessing unit, can boost uranium utilization to over 95\% while shortening the half-life of highly toxic waste from hundreds of thousands of years to about 500 years and reducing its volume by 96\%. To realize this concept, a high-power superconducting linac drives a spallation target to produce neutrons for a subcritical fast reactor, and the spent fuel is regenerated through a low-contamination dry process, enabling continuous recycling without deep purification. The first megawatt-scale experimental facility, the China initiative Accelerator Driven System (CiADS), is under construction in Huizhou, serving as the world's first integrated ADS for testing this transformative technology \cite{He:2023izb,WangRuoXu:2024nst,DENG2024110202,CHEN2026104246}.

To achieve a high-stability, high-reliability, high-safety, and controllable nuclear energy system, a distinctive feature of ADANES is its strong emphasis on automation technology and artificial intelligence (AI). The newly launched ``AI for ADANES'' initiative embeds a physics-principle-constrained AI system into the entire life cycle -- from design and commissioning to operation and maintenance \cite{LiMenghan2026AIRoadmap}. A new paradigm driven by operational data, physical models, and accumulated expert knowledge is constructed to reconcile AI's ``black-box'' nature with the nuclear industry's zero-tolerance safety requirements. High-fidelity multiphysics simulations serve as independent inputs to teach and constrain the AI-assisted control system of ADANES. Accurate, high-precision nuclear reaction models are a crucial basis for the simulation software.

The reliable spallation reaction model, well constrained by experimental data, is of particular importance, as it is not realistic to acquire all detailed experimental data of all nuclei over a broad kinematical region to create a complete nuclear data library at high incident energy. The nuclear spallation model still plays a crucial role in high-energy Monte-Carlo simulations. Spallation is an intense and complex reaction process in which a projectile particle strikes a heavy atomic nucleus, generating numerous lighter particles -- such as neutrons, protons, and small nuclear fragments -- and leaving a residual nucleus that is significantly lighter than the original \cite{David:2015ura}. The projectile energy spans from tens to thousands of MeV. The spallation reaction is commonly described within a two-stage framework. The first stage, the intranuclear cascade, is a fast process ($10^{-22}$ s) in which the incident particle initializes a series of binary collisions between the nucleons inside the target nucleus, ejecting high-energy particles and leaving the residual nucleus in a highly excited state. A number of models have been developed to simulate the first stage, such as the semiclassical codes ISABEL \cite{Yariv:1979zz,Yariv:1981vb}, INCL \cite{Boudard:2012wc,Mancusi:2014fba}, and CRISP \cite{Deppman:2004vc}, and the quantum molecular dynamics (QMD) model \cite{Bass:1998ca,Feng:2005anh,Ou:2008zza,Su:2019mrl}. The second stage, de-excitation, is a relatively slow process ($10^{-16}$ s) in which the remnant nucleus de-excites through nucleon evaporation, fission, and other decay channels. Popular de-excitation packages include the binary decay code GEMINI \cite{Charity:1988zz,Charity:2010wk,Mancusi:2010tg}, the evaporation-fission code ABLA \cite{Junghans:1998vzp,Kelic:2009yg}, and the SMM code based on the statistical model \cite{Bondorf:1995ua,Botvina:2000jc}. Recently, the constrained molecular dynamics model has provided the first complete dynamical description of the spallation process without artificially separating the two stages \cite{Assimakopoulou:2018nsj}. Nevertheless, the computing time required to simulate the slow de-excitation stage using the molecular dynamics model is prohibitively long.

Despite the successes of physics-based models of intranuclear cascade and residue de-excitation, current microscopic dynamical models of nuclear spallation face several challenges. Some evident deviations between models and data are observed \cite{David:2015ura,Iwamoto:2020epjconf}, and the corresponding bias and uncertainty of model predictions are not yet fully characterized \cite{Schnabel:2018yft,Hirtz:2026cgr}. First, no model can quantitatively describe all reaction channels well at the same time. A model that precisely reproduces nuclear fragment cross sections probably cannot describe the differential cross sections of nucleon or light-cluster emission well. The dynamics of spallation reactions is very complex because the projectile energy covers several orders of magnitude and the nuclear targets differ greatly. Second, the details of differential cross sections and the nuclear dependence are not well described by physics-based dynamical models \cite{Schnabel:2018yft}. Although the total cross sections of nucleon emission are usually well reproduced by the models, there are noticeable discrepancies between the dynamical reaction models and the experimental data in terms of the energy spectra of emitted nucleons. An accurate neutron energy spectrum is critical for assessing the power of an ADS burner \cite{WeiXiaoQiang:2026nst}. Third, it is not easy to optimize the parameters of the dynamical reaction models \cite{Hirtz:2026cgr,Hirtz:2023wfc}, and the model parameters are usually not comprehensively tuned. Fourth, different physics-based models predict quite different results \cite{David:2015ura,WeiXiaoQiang:2026nst}. Sometimes, the difference between models is larger than the relevant experimental uncertainties. Last but not least, some underlying reaction mechanisms are not implemented in the dynamical models, such as nucleon-nucleon short-range correlations \cite{Rodriguez-Sanchez:2024bjn}, nuclear medium effects \cite{Aichelin:1991xy,Li:1993rwa,Li:1993ef,Li:2008gp}, and the proper treatment of the nuclear Fermi surface \cite{Mancusi:2014eia}.

The recent booming development of data science and the extensive successful applications of AI inspire us to construct a data-based model for the precise description of nuclear reactions and to explore machine-learning techniques for tackling the complex nuclear and energy dependencies of differential cross sections. A data-based artificial neural network (ANN) model is effective and efficient in reproducing the vast amount of experimental data. The machine-learning method of ANN is highly flexible and unbiased in describing complicated multi-variable functions, offering a powerful tool for making predictions once the ANN model is properly trained with adequate data. In fact, several studies have been published on applying Bayesian neural network (BNN) to describe fragment production cross sections \cite{Ma:2020mbd,Peng:2021tzc,Ma:2022dwb,Song:2022yjr}, achieving much smaller deviations compared with experimental data. It has also been found that physics-guided BNN models give more reasonable extrapolation results \cite{Peng:2021tzc,Ma:2022dwb}. Some analyses using BNN to reproduce the total cross sections of specific channels of spallation reactions have been performed, and the energy dependence of the cross section can be described by BNN in a certain energy range \cite{Peng:2021tzc}. Recently, high-energy neutron-induced fission cross sections have been successfully modeled with the BNN method \cite{Zhang:2026nst}. Moreover, the residual errors between the QMD model and experimental data have been precisely and effectively modeled with BNN \cite{Song:2022yjr}. Some researchers have even applied an ANN framework to learn simulation data from the INCL+ABLA model in order to achieve accurate predictions of fission-yield cross sections \cite{Rodriguez-Sanchez:2025paj}.

Precise descriptions of neutron energy spectra and production cross sections in spallation reactions are indispensable for the design and simulation of ADS. High-energy neutrons induce more fission events and more neutron emissions in an ADS reactor, resulting in a higher value for nuclear transmutation compared with low-energy neutrons. In ADS, the neutron spallation targets are properly located inside the subcritical reactor. There is no doubt that precise descriptions of neutron energy and angular distributions are essential. In this work, we propose to construct a novel data-based reaction model within an ANN framework to describe the neutron double-differential cross sections (DDCS) of nuclear spallation. In fact, a decent number of neutron DDCS data for spallation reactions from many groups are available. Therefore, it is quite interesting to see whether a general ANN model can reproduce the experimental data precisely. Moreover, it is also meaningful to verify the predicted dependencies of neutron DDCS on projectile energy, nuclear target, neutron angle, and neutron energy from the purely data-driven machine-learning model. In addition, it is interesting to identify the key input variables that determine the neutron DDCS based on the ANN model analysis. So far, no study has applied an ANN framework to study the double-differential cross sections of nuclear spallation. Hence, this paper presents, for the first time, the development of a purely data-based ANN framework to precisely describe the neutron DDCS in spallation reactions. In addition to building an ANN reaction model of DDCS, we also pay attention to testing the predictive power and generalization ability of the obtained ANN model.

The organization of the paper is as follows. The neutron DDCS measurements of nuclear spallation reactions are reviewed in Sec. \ref{sec:experimental-data}. The experimental data for training and testing the ANN model are collected and listed in Sec. \ref{sec:experimental-data} as well. The proposed ANN framework and the training method are detailed in Sec. \ref{sec:ann-framework}. The outcomes of different regularization methods and different input variables are shown and discussed in Sec. \ref{sec:results}. Comparisons between the final ANN model predictions and the experimental data are also presented in Sec. \ref{sec:results}. A brief summary of this study and an outlook for the ANN model in nuclear reaction studies are given in Sec. \ref{sec:summary}.

\section{Experimental data sets}\label{sec:experimental-data}

A pertinent feature of the ANN model is that it relies heavily on experimental data to optimize the model. Fortunately, sufficient measurements of the DDCS of neutron production from various nuclear targets over a broad range of projectile energies are available. Many accelerator facilities worldwide can deliver proton beams of intermediate to high energies. The neutron DDCS in spallation reactions were measured using the time-of-flight (TOF) technique, with which neutron energy spectra can be measured with satisfactory energy resolution over a wide energy range. Particular efforts were made to discriminate and suppress gamma-ray background contributions. Gamma-ray contamination was usually handled by interposing gamma-ray filters to attenuate the high flux, setting thresholds, discriminating the pulse shapes and durations of scintillation signals, applying TOF cuts, and/or performing offline data corrections. Note that thin targets were used in some experiments, whereas thick targets were chosen in others to increase the statistics of spallation neutrons. The multiple-scattering effect in a thick target was quantitatively checked through calculations with several widely used Monte-Carlo codes. The thick-target correction is made or controlled at an acceptable level. We selected a set of mandatory experimental data for training the ANN model and reserved some additional data for testing the model.

\subsection{Selected data for training ANN model}
\label{subsec:data-for-training}

\begin{table*}[htp]
\centering
\caption{
Experimental data set of neutron DDCS used for training the ANN model for spallation reactions. The proton beam energies, neutron emission angles, and references for the selected data are listed.
}
\label{tab:train-data}
\begin{tabular}{cccc}
\hline\hline
Target  & Beam energy [MeV]   &   Neutron angle [Deg.]                        &  Reference                 \\
\hline
Al      &  256/800/1200       &   0/7.5/10/25/30/40/55/60/85/100/110/120/130/145/150/160  & \cite{Meier:1992anx,Amian:1992jal,Leray:2001pp}  \\
Fe      &  256/800/1200/1600  &   0/7.5/10/25/30/40/55/60/85/100/115/120/130/145/150/160  & \cite{Meier:1992anx,Amian:1992jal,Leray:2001pp}  \\
Zr      &  1200               &   0/10/25/40/55/85/100/115/130/145/160                    & \cite{Leray:2001pp}  \\
Cd      &  800                &   30/60/120/150                                           & \cite{Amian:1992jal}  \\
W       &  800/1200           &   0/10/25/30/40/55/60/85/100/115/120/130/145/150/160      & \cite{Amian:1992jal,Leray:2001pp}  \\
Pb      &  62.9/256/800/1200/1600 & 0/7.5/10/24/25/30/35/40/55/60/70/80/85/100/115/120/130/145/150/160  & \cite{Meier:1992anx,Amian:1992jal,Leray:2001pp,Guertin:2004}  \\
Th      &  1200               &   0/10/25/55/85/130/145/160                               & \cite{Leray:2001pp}  \\
U       &  256/800            &   7.5/30/60/120/150                                       & \cite{Meier:1992anx,Amian:1992jal} \\
\hline\hline
\end{tabular}
\end{table*}

As an initial study of an ANN reaction model, we focus on spallation induced by high-energy protons, because experimental data for proton-induced spallation reactions are much more abundant and cover a broad range of projectile energies. Since we are more interested in heavy nuclei, we collect only experimental data with mass numbers above 27 (including $^{27}$Al). In this work, we use the mandatory data for the benchmark of spallation models organized by the International Atomic Energy Agency in 2010 \cite{Takeda:2011vmy,web_Benchmark_IAEA2010}, to train a commonly used ANN framework for describing neutron DDCS. Table \ref{tab:train-data} summarizes the experimental data we selected to train the ANN reaction model \cite{Meier:1992anx,Amian:1992jal,Leray:2001pp,Guertin:2004}. The nuclear targets include aluminum, iron, zirconium, cadmium, tungsten, lead, thorium, and uranium, eight nuclides in total. Iron, lead, and depleted uranium are typical structural, spallation-target, and ADS-fuel materials, respectively. The lowest projectile energy in the data is 62.9 MeV, and the highest is 1600 MeV. The measured neutron emission angles include 0, 7.5, 10, 24, 25, 30, 35, 40, 55, 60, 70, 80, 85, 100, 110, 115, 120, 130, 145, 150, and 160 degrees, with forward and backward angles well covered. These data were measured by different collaborations at several accelerator facilities: LAMPF at Los Alamos, USA \cite{Meier:1992anx,Amian:1992jal}; SATURNE at Saclay, France \cite{Leray:2001pp}; and CYCLONE in Belgium \cite{Guertin:2004}. Thus, the analysis using the ANN reaction model in this work is not only an exploration of a data-driven reaction model but also an assessment of the consistency among measurements from different groups worldwide. 

The total number of collected training data points is 8310, which is neither very small nor excessively large. These data certainly contain some clear patterns of the underlying physical mechanisms of spallation reactions. We hope that some fundamental laws of nuclear spallation reactions can be learned by the machine-learning technique of the ANN framework. If the ANN model is constructed and trained carefully, the dependencies of the spallation DDCS on nuclear target, incident particle energy, neutron energy, and angle can be captured.

To incorporate the uncertainties of the measured DDCS data, we generate multiple replicas of the experimental data for ANN model training. The replica data set is generated according to Gaussian distributions based on the statistical uncertainties of the measurements. The ANN model is then trained on the replica data instead of the original experimental data, so that the noise in the experimental data acts as a kind of weight in optimizing the ANN model. More details about the data replicas can be found in subsection \ref{subsec:data-replica}. 

\subsection{Data reserved for testing ANN model}
\label{subsec:data-for-testing}

\begin{table}[htp]
\centering
\caption{
Reserved experimental neutron DDCS data used only for testing the final trained ANN model (i.e., not used for training the ANN model).
}
\label{tab:untrain-data}
\begin{tabular}{cccc}
\hline\hline
Target   &   Beam energy [MeV]     &   Neutron angle [Deg.]      &  Reference \\
\hline
Fe       &      597                &    30/60/120/150            & \cite{Amian:1993dhv}  \\
Pb       &      1000               &    15/60/90/120             & \cite{Trebukhovsky:2003uh}  \\
In       &      1500/3000          &    15/30/60/90/150          & \cite{Ishibashi:1997gbe}  \\
\hline\hline
\end{tabular}
\end{table} 

To independently test the trained ANN model, we reserve some experimental data solely for final model testing \cite{Amian:1993dhv,Trebukhovsky:2003uh,Ishibashi:1997gbe}. The data reserved solely for testing the ANN model are listed in Table \ref{tab:untrain-data}. The beam energies (597, 1000, 1500, and 3000 MeV) of the test data differ from those of the training data, and 3000 MeV lies well outside the energy range covered by the training data (62.9 - 1600 MeV). Hence, these test data are useful for checking the energy dependence of the DDCS predicted by the ANN model and for assessing the extrapolation power of the model. The indium target (In, $Z = 49$) is not included in the training data but only in the test data. Thus, the test data for the In target are valuable for testing the nuclear dependence of the DDCS predicted by the ANN model. A neutron emission angle of $90^\circ$ is not covered by the training data. Therefore, the DDCS data at $90^\circ$ are important for validating the angular dependence of the DDCS given by the trained ANN model. The reserved experimental data set for model testing is not only crucial for an independent test of the ANN model but also necessary to evaluate the generalization ability of the model. 

\subsection{Uncertainty-incorporated replicas of experimental data}
\label{subsec:data-replica}

Every experimental measurement provides the corresponding uncertainties of the data, and the uncertainties are as important as the measured central values. To incorporate the uncertainty information of the data into the machine-learning approach of the ANN framework and to suppress overfitting caused by using only the central values of the data, we generate replicas of the experimental data that contain the data uncertainty information. Generating data replicas is a simple approach in this work. A random Gaussian error is added to the measured value, as
\begin{equation}
\begin{split}
   y^{\rm replica}=y^{\rm exp.}+\delta,\qquad \delta \sim \mathcal{N}(0,\sigma^2).
\end{split}
\label{eq:replica-generation}
\end{equation}
Currently, only the statistical uncertainty $\sigma^{\rm stat.}$ is included when generating the data replicas. In this way, the data replicas are randomly distributed around the measured central values, containing the information of statistical uncertainties.

To control fluctuations arising from the small number of samplings when randomly generating data replicas by Gaussian smearing, we generate 30 replicas for each data point in this study. In total, we obtain 249,300 replica data points for training the ANN model of neutron DDCS. The uncertainty in the bin size of the neutron energy spectrum is not incorporated into the data replicas, since the bin size is not large. Only smearing of the measured DDCS according to the experimental uncertainty is implemented for the data replicas.

\section{ANN framework for precision DDCS prediction}\label{sec:ann-framework}

\begin{figure*}[htbp]
\begin{center}
\includegraphics[width=0.7\textwidth]{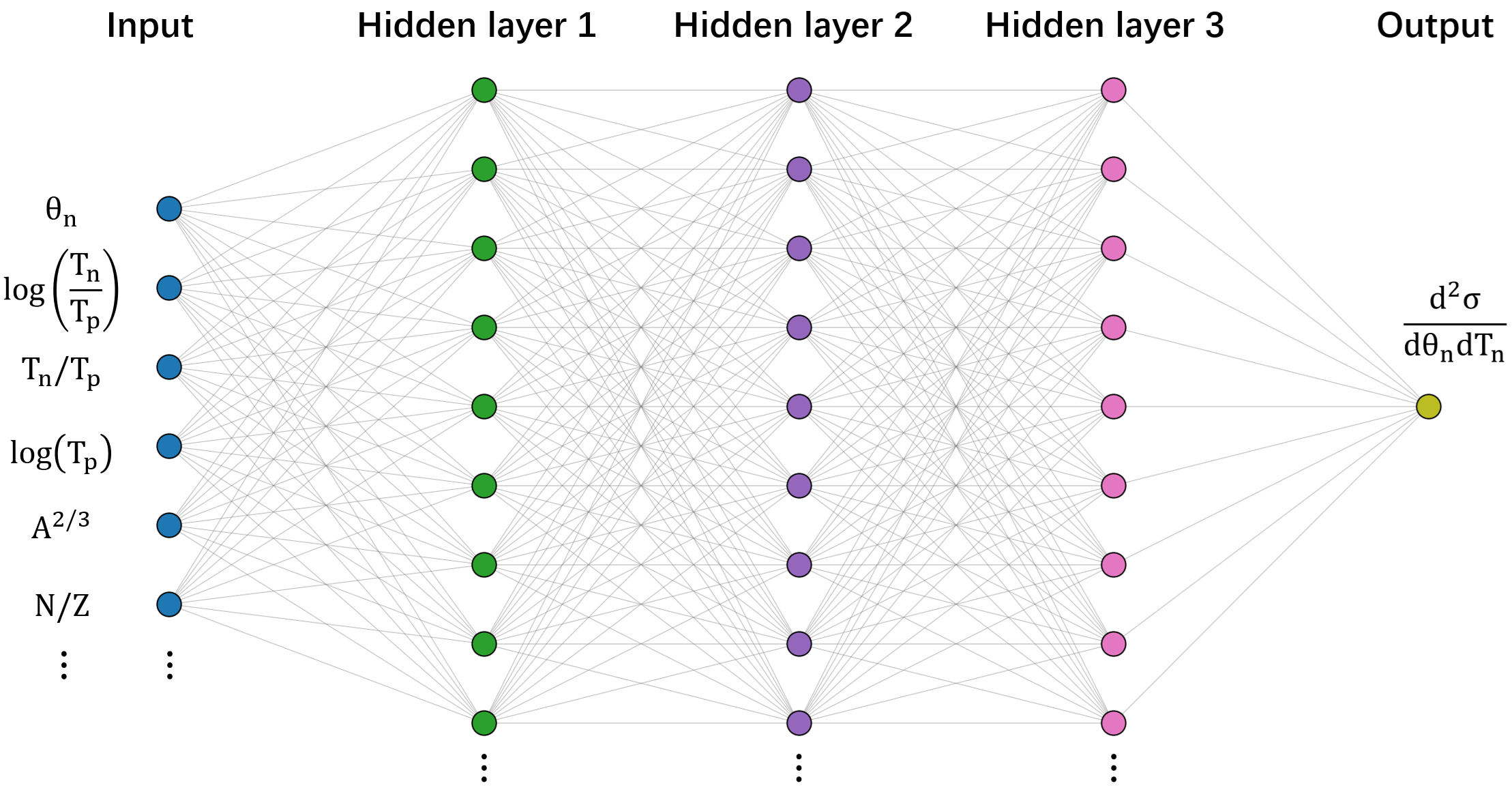}
\caption{(Color online) An schematic diagram of the three-layer artificial neural network adopted in this study for the machine learning of neutron DDCS in nuclear spallation reaction.  
}
\label{fig:ANN-framework-illustration}
\end{center}
\end{figure*}

In this work, we adopt a multilayer perceptron (MLP) network \cite{rumelhart1986learning,rosenblatt1958perceptron} to perform regression analysis of the neutron DDCS data in terms of a set of input variables. The MLP network is also known as a ``back-propagation'' or ``feedforward'' network, and it is the most widely used type of ANN. Fig. \ref{fig:ANN-framework-illustration} illustrates an example of an MLP network for predicting the neutron DDCS in nuclear spallation, with three hidden layers of neurons. The upper hidden layer is fully connected to the lower hidden layer, and each neuron in the hidden layers has a nonlinear activation function of the input, enabling the extraction of nonlinear features of the nuclear reaction cross section. Let the upper and lower hidden layers be denoted as ${\bf X}=\{X_i\}$ and ${\bf Y}=\{Y_i\}$, respectively; then the output of a neuron in the lower hidden layer is computed as
\begin{equation}
\begin{split}
   Y_j = a_j + {\rm activation} \left( \sum_{i=1}^{I} b_{ji} X_i \right),
\end{split}
\label{eq:neural-output}
\end{equation}
where $b_{ji}$ is the connection weight between $X_i$ and $Y_j$, $I$ is the total number of neurons in the upper hidden layer, $a_j$ is the bias parameter for the output $Y_j$, and ${\rm activation}$ denotes the activation function of the neuron. These weights and biases ${\bf \theta}=\{a_i, a_j, b_{ji}, \ldots\}$ are the trainable parameters of the network that can be optimized with a gradient descent algorithm. ReLU and Softplus activation functions are commonly used for regression problems. For the neuron in the output layer, the activation function is simply a linear function that generates a value from the inputs of the upstream layer.

The input of the ANN should contain the physical variables that are relevant to the underlying spallation mechanism or that determine the neutron DDCS. Our main purpose in this work is to build an ANN model for a precise description of the neutron DDCS in proton-induced spallation reactions, a reaction written as $p+A\rightarrow xn+X$. Therefore, the input variables should include at least the type of nuclear target, the kinematics of the projectile, and the kinematics of the emitted neutron. Fig. \ref{fig:ANN-framework-illustration} shows an example of the ANN input variables. 

The sole output node of the ANN is designed to predict the DDCS of neutron emission, a key quantity of interest that is inherently nonlinear and sensitive to the interplay of the input variables.

\subsection{ANN input variables}
\label{subsec:ann-input}

The input layer of the ANN model for spallation cross sections is straightforward; it should include all variables that describe the initial state of the $p$-$A$ collision. These include the projectile particle type, the target type, and the incident energy. Since we focus on proton-induced spallation reactions, the minimum input variables are the projectile energy $T_p$, the atomic number $Z$ and mass number $A$ of the target which uniquely define the target nucleus. The neutron DDCS are functions of the neutron emission angle $\theta_n$ and neutron energy $T_n$, which are also important input variables for the model. 

As we use a small-scale ANN for the DDCS regression study, more relevant input variables can accelerate the convergence of the ANN model during optimization. Considering the possible physical mechanisms of nuclear reactions, we propose the following options for the input variables of the ANN: $\theta_n$, $1/\sin^4(\theta_n)$, $\log(T_n/T_p)$, $(T_n/T_p)$, $\log(T_p)$, $T_p$, $A^{2/3}$, and $N/Z$. In addition to the normal angular variable, $1/\sin^4(\theta_n)$ is proposed for the possible angular dependence of the quasi-elastic process, taken in the form of the Rutherford scattering cross section. The quasi-elastic peak decreases quickly with increasing scattering angle $\theta_n$. To obtain a universal pattern for the neutron spectrum that is independent of the projectile energy, we suggest using the dimensionless neutron energy variables $\log(T_n/T_p)$ and $(T_n/T_p)$. Since the emitted neutron energy covers several orders of magnitude, the logarithm of the dimensionless neutron energy would be a good choice for describing the power-law behavior of the neutron energy spectrum. Besides the regular projectile energy $T_p$, the logarithm $\log(T_p)$ is also suggested to describe the slow variation of the DDCS with projectile energy at high-energy scales. Because the nuclear radius approximately scales as $A^{1/3}$, $A^{2/3}$ would be a good quantity for describing the geometrical cross section of the nuclear target. Therefore, we take $A^{2/3}$ instead of the simple mass number $A$. Another important property of the target is the neutron-proton asymmetry or isospin of the target, which can be well quantified by the ratio $N/Z$.

The outlined eight optional variables of the ANN input layer are important for high-precision modeling of the neutron DDCS data. Nevertheless, some kinematic variables may be redundant in determining the neutron DDCS in spallation reactions. Whether the suggested input variables are vital and of key importance to the ANN model should be tested through detailed analysis during ANN training. If an input variable is omitted without influencing the agreement between the ANN model and the experimental data, then this input variable is not essential for the ANN model. The question of which input variables are key to the ANN model of neutron DDCS is studied and discussed in Sec. \ref{sec:results}.

\subsection{ANN model configuration}
\label{subsec:ann-construction} 

In this study, we employ a small-scale ANN for describing the neutron DDCS of nuclear spallation reactions. The input and output layers are discussed in the previous subsections. Regarding the hidden layers, the current ANN version consists of three hidden layers, each containing 128 neurons. To introduce the necessary nonlinearity while maintaining smooth gradients for stable backpropagation, we employ the softplus activation function \cite{dugas2000incorporating}, defined as $f(x)=\ln(1+e^{x})$, in all hidden layers. Unlike the rectified linear unit (ReLU), softplus provides a differentiable, non-sparse approximation of the linear rectifier, which helps to avoid dead-neuron issues and ensures a well-behaved derivative across the entire real domain -- a desirable property for regression tasks where output continuity is critical. 

Each neuron has a bias parameter, and each connection between two neurons has a free weight parameter. Thus, the total number of trainable parameters for the proposed ANN model is 34,305 in the case of eight input variables. The number of trainable parameters is not large for modern computers, yet it is sufficient to model a complicated function with fewer than ten input dimensions. How to optimize the free parameters and train the ANN model is discussed in the following subsection.

\subsection{ANN training strategy}
\label{subsec:ann-training-method} 

To train the ANN model, replicas of the experimental data are used instead of directly using the central values of the data. The generated replica data are split equally into a training data set and a validation data set; the training set is used for ANN training, and the validation set is used to check whether overfitting occurs. The entire training process is monitored using a validation set, and appropriate early stopping is applied to prevent over-training. Before splitting, the replica data are shuffled randomly to eliminate bias due to data splitting. The free parameters of the ANN are updated by comparing predictions with each small batch of data. The batch size is 64, which is beneficial for escaping local minima of the loss function, avoiding overfitting, and ensuring good generalization ability. A balance between gradient noise and computational efficiency is maintained. The total number of epochs for ANN training is finally chosen to be 300 by evaluating the training and validation loss curves.

The loss function for evaluating model performance is taken as the mean squared error (MSE), defined as ${\rm MSE}=\frac{1}{n}\sum_{i=1}^{n}(y_i-\hat{y}_i)^2$, where $y_i$ denotes the data and $\hat{y}_i$ denotes the model prediction. The MSE is most appropriate for continuous numerical prediction (regression) tasks, and the gradient of the MSE is proportional to the error itself, which provides smooth and consistent updates during backpropagation. We apply the Adam (adaptive moment estimation) algorithm \cite{kingma2015adam,reddi2018convergence} to optimize the ANN parameters. The Adam method computes individual adaptive learning rates for each parameter by estimating first- and second-order moments of the gradients, thus providing efficient and robust stochastic optimization. In this study, the learning rate itself is not fixed but is adaptively tuned per iteration through the Adam update rules, with the initial learning rate set relatively large (0.015) to allow the algorithm to self-adjust finely based on the gradient history. This adaptive scheme has proven particularly effective in navigating the complex loss landscape characteristic of high-dimensional regression problems.

To overcome possible overparameterization or overfitting, we comprehensively explore two typical regularization strategies: L2 weight decay \cite{krogh1992simple} and dropout \cite{srivastava2014dropout,hinton2012improving}. In the L2 approach, a penalty term proportional to the squared Euclidean norm of the weight matrices is added to the loss function, encouraging the network to maintain small weights and thus reducing the effective capacity of the model. In the dropout scheme, a randomly selected subset of neurons (with a prescribed dropout probability) is temporarily ``dropped out'' (i.e., set to zero) during each forward pass, forcing the network to learn redundant representations and preventing co-adaptation of hidden units. Both regularization techniques were evaluated separately and in combination. The effects of regularization on model prediction precision and overfitting are examined in Sec. \ref{sec:results}.

\section{Results and discussion}\label{sec:results}

In this section, we present, through a series of comparative studies, the influence of regularization schemes and the final choice for ANN training, the optimal set of key input variables for the ANN framework, the goodness of fit of the ANN model to the training data, the extrapolation power of the ANN model evaluated using test (unseen) data, and the ANN predictions of the neutron DDCS for proton-induced spallation reactions on a copper target at various proton energies.

\subsection{Regularization strategy for ANN training}
\label{subsec:regularization-study}

\begin{figure}[htbp]
\begin{center}
\includegraphics[width=0.43\textwidth]{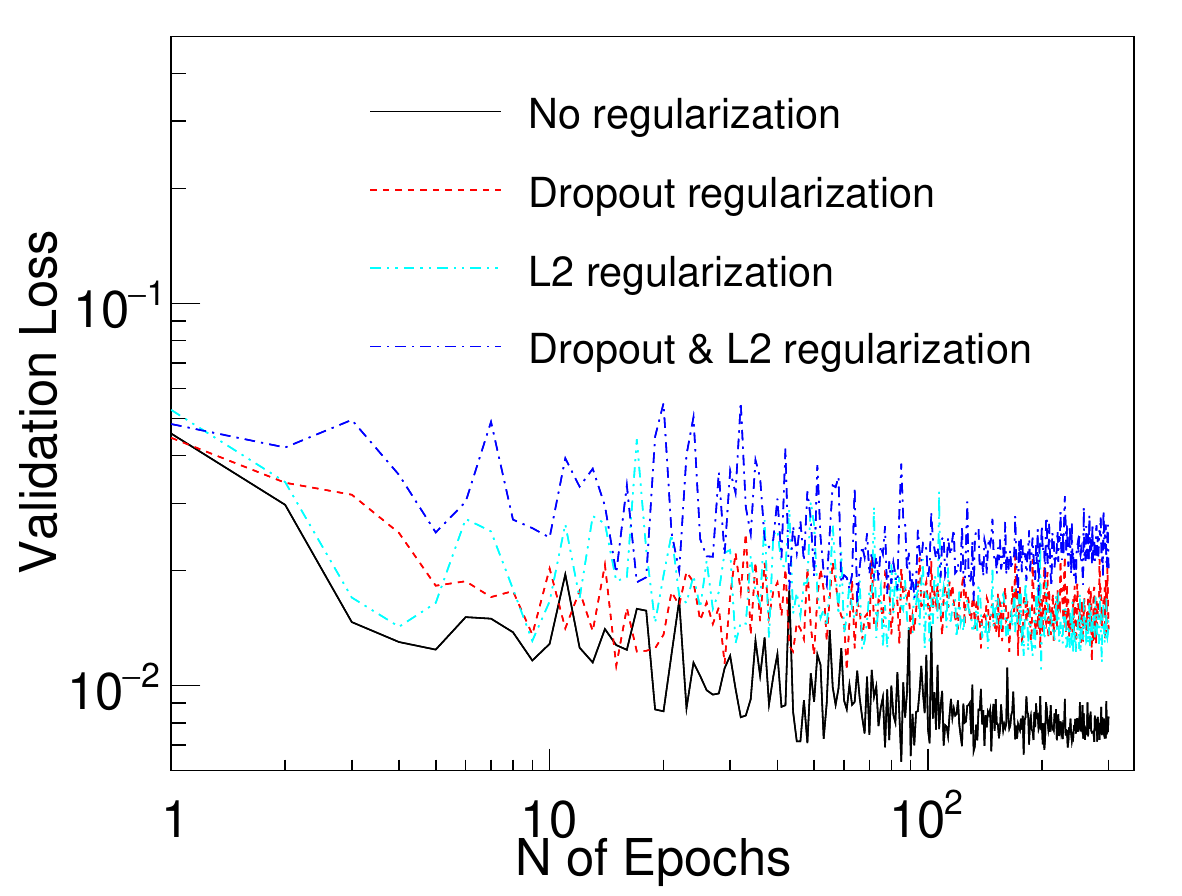}
\caption{
(Color online) Validation loss functions as a function of the number of training epochs under different regularization schemes: without regularization (black solid curve), with dropout regularization (red dashed curve), with L2 regularization (cyan dash-dot-dot curve), and with both dropout and L2 regularizations (blue dash-dotted curve).
}
\label{fig:Validation_losses_with_diff_regularizations}
\end{center}
\end{figure}

\begin{figure}[htbp]
\begin{center}
\includegraphics[width=0.43\textwidth]{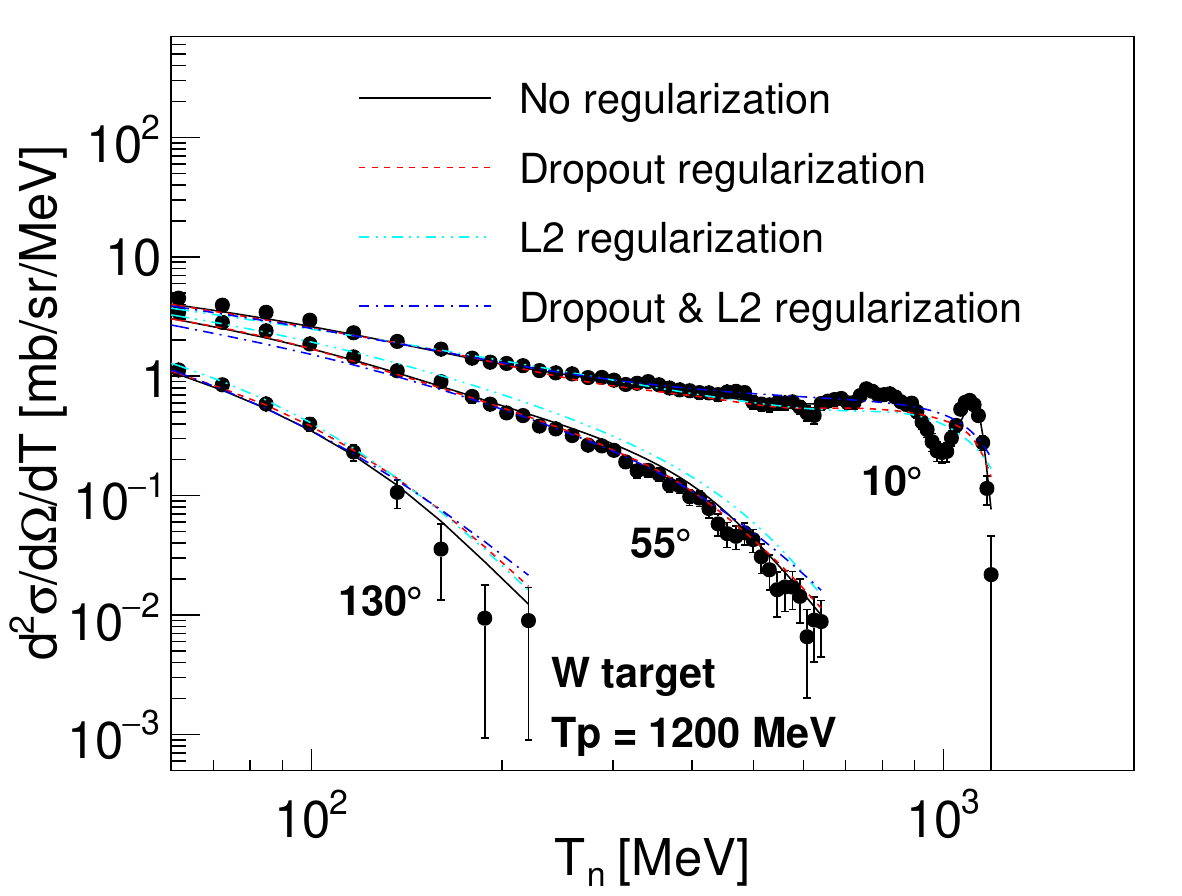}
\caption{
(Color online) ANN predictions for the neutron DDCS as a function of the emitted spallation neutron energy under different regularization schemes: without regularization (black solid curve), with dropout regularization (red dashed curve), with L2 regularization (cyan dash-dot-dot curve), and with both dropout and L2 regularizations (blue dash-dotted curve). The ANN predictions are for the proton-induced spallation reaction on a W target at 1200 MeV, and they are compared with the experimental data \cite{Leray:2001pp}.
}
\label{fig:W_1200MeV_diffAngles_diffReguls}
\end{center}
\end{figure}

Fig. \ref{fig:Validation_losses_with_diff_regularizations} displays the validation losses as a function of the number of epochs (iterations) under different regularization schemes and without regularization. All loss functions begin to stabilize after about 200 epochs. ANN training without regularization yields the smallest validation loss, whereas training with both L2 and dropout regularization gives the largest validation loss. The stabilized loss value is similar for training with either L2 regularization or dropout regularization, lying between the value obtained without regularization and that obtained with both regularization schemes. A larger loss value indicates larger deviations between the trained ANN model and the experimental data.

Fig. \ref{fig:W_1200MeV_diffAngles_diffReguls} shows the predicted neutron DDCS from ANN models trained under different regularization schemes and without regularization, compared with experimental measurements of high-energy proton-induced spallation reactions on a tungsten target. All ANN predictions of the neutron energy spectra are monotonic and smooth when regularization is used. The quasi-elastic and quasi-inelastic peaks in the neutron energy spectra at small angles are not well reproduced by the ANN model with either dropout or L2 regularization. The ANN model trained with both L2 and dropout regularization deviates most from the experimental data. Only the ANN model without regularization reproduces the peaks in the neutron energy spectra, yielding the best consistency with the data. The predicted shapes of the neutron energy spectra agree with the loss-function results discussed above.

\begin{figure}[htbp]
\begin{center}
\includegraphics[width=0.43\textwidth]{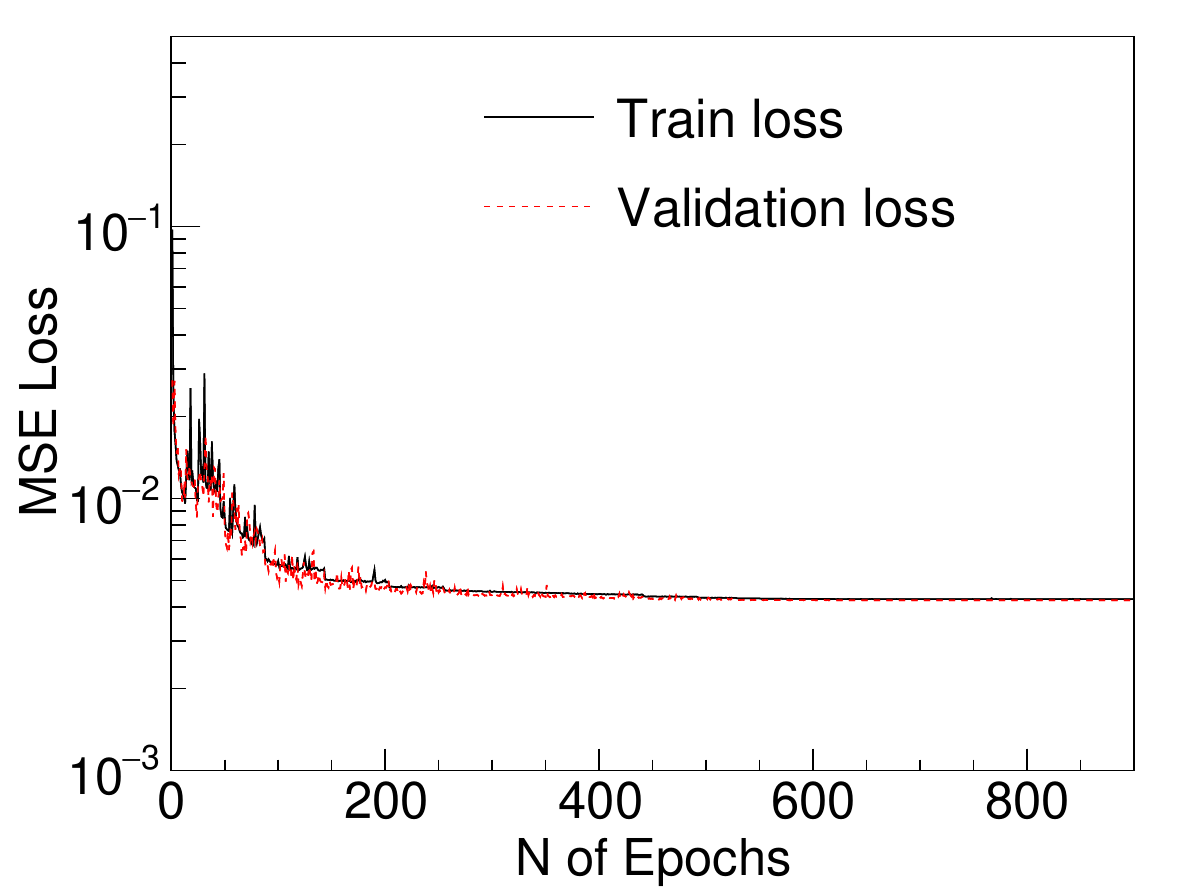}
\caption{
(Color online) Evolution of the training and validation losses with the number of epochs for ANN training.
}
\label{fig:Train_and_validation_losses}
\end{center}
\end{figure}

Since the ANN trained without any regularization can describe the quasi-elastic and quasi-inelastic peaks well, we decide not to apply regularization in the final ANN setup for learning the neutron DDCS data of spallation reactions. In Fig. \ref{fig:W_1200MeV_diffAngles_diffReguls}, it can be seen that small fluctuations in the experimental data are not fitted by the ANN model even without regularization. The overfitting problem is inherently avoided by using a large sample of replica data that contains the noise from experimental measurements. A typical and definitive sign of overfitting is that the validation loss starts to increase while the training loss continues to decrease or stabilize. The training and validation losses are the loss functions computed with the training data set and the validation data set, respectively. Fig. \ref{fig:Train_and_validation_losses} presents the training and validation losses evolving over a large number of training iterations. The training and validation losses stabilize after about 300 epochs, and no increase in the validation loss is observed even after 900 epochs. Hence, no overfitting of the ANN model is found. Using a large sample of uncertainty-incorporated replicas of the experimental data prevents overfitting of the ANN model.

The benefit of training the ANN with noise-containing replicas of the experimental data is twofold. First, experimental error information is included in optimizing the ANN model when replica data with Gaussian error smearing are used. Thus, the variation in the loss function has different sensitivities to experimental data with different errors. Second, the noise in the replica data hinders over-optimization of the ANN model. To ensure that the replica data fully reflect the experimental errors, a big number of replicas for each data point are required. The total number of replica data is 249,300, and the number of replica data for ANN training is 124,650, which is much larger than the number of free parameters in our proposed small-scale ANN model (34,305). Using a small-scale ANN is another reason for the absence of overfitting without applying any regularization.

\subsection{Key input variables for ANN model}
\label{subsec:input-variable-study}

\begin{figure}[htbp]
\begin{center}
\includegraphics[width=0.43\textwidth]{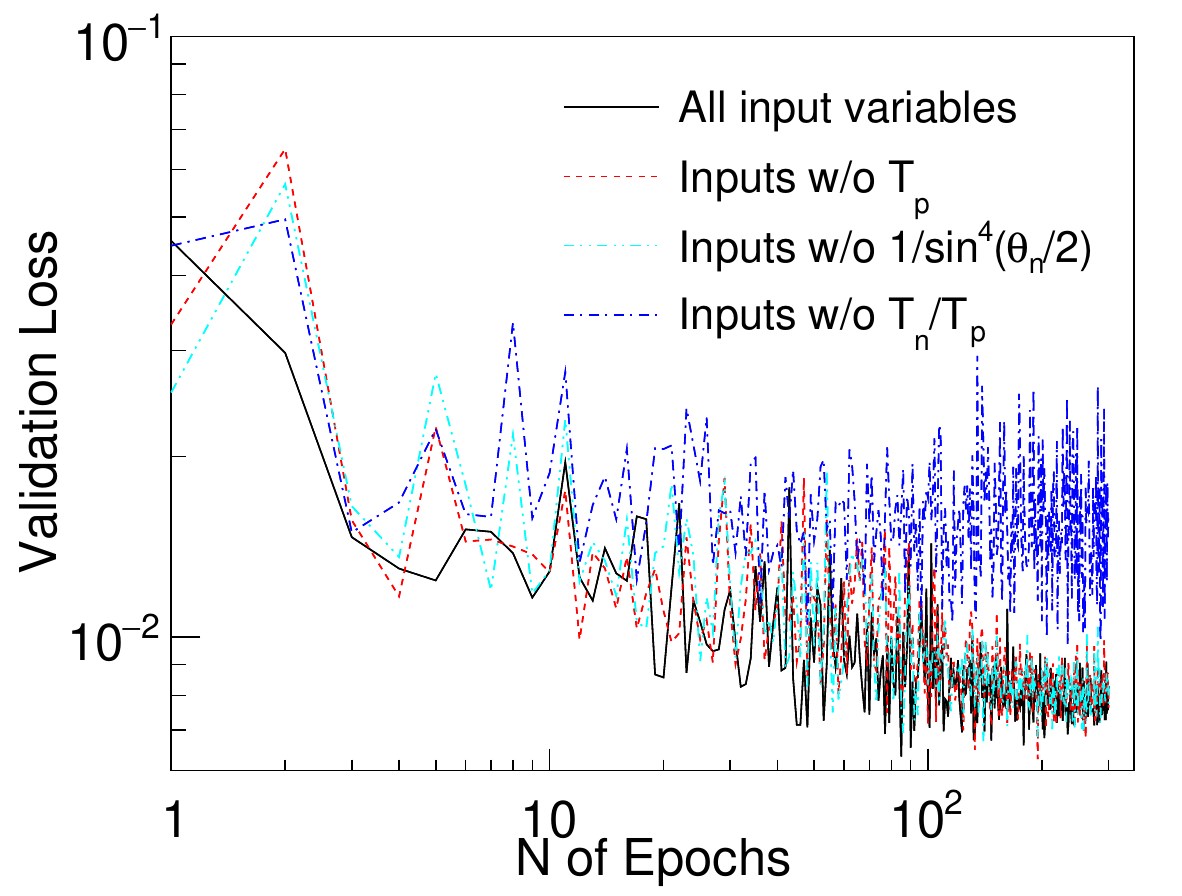}
\caption{
(Color online) Validation loss functions as a function of the number of training epochs using different input variables for the ANN framework: all input variables $\{\theta_n, 1/\sin^4(\theta_n), \log(T_n/T_p), (T_n/T_p), \log(T_p), T_p, A^{2/3}, N/Z\}$ (black solid curve), with only the variable $T_p$ removed (red dashed curve), with only the variable $1/\sin^4(\theta_n)$ removed (cyan dash-dot-dot curve), and with only the variable $T_n/T_p$ removed (blue dash-dotted curve).
}
\label{fig:Validation_losses_with_diff_input_variables}
\end{center}
\end{figure}

\begin{figure}[htbp]
\begin{center}
\includegraphics[width=0.43\textwidth]{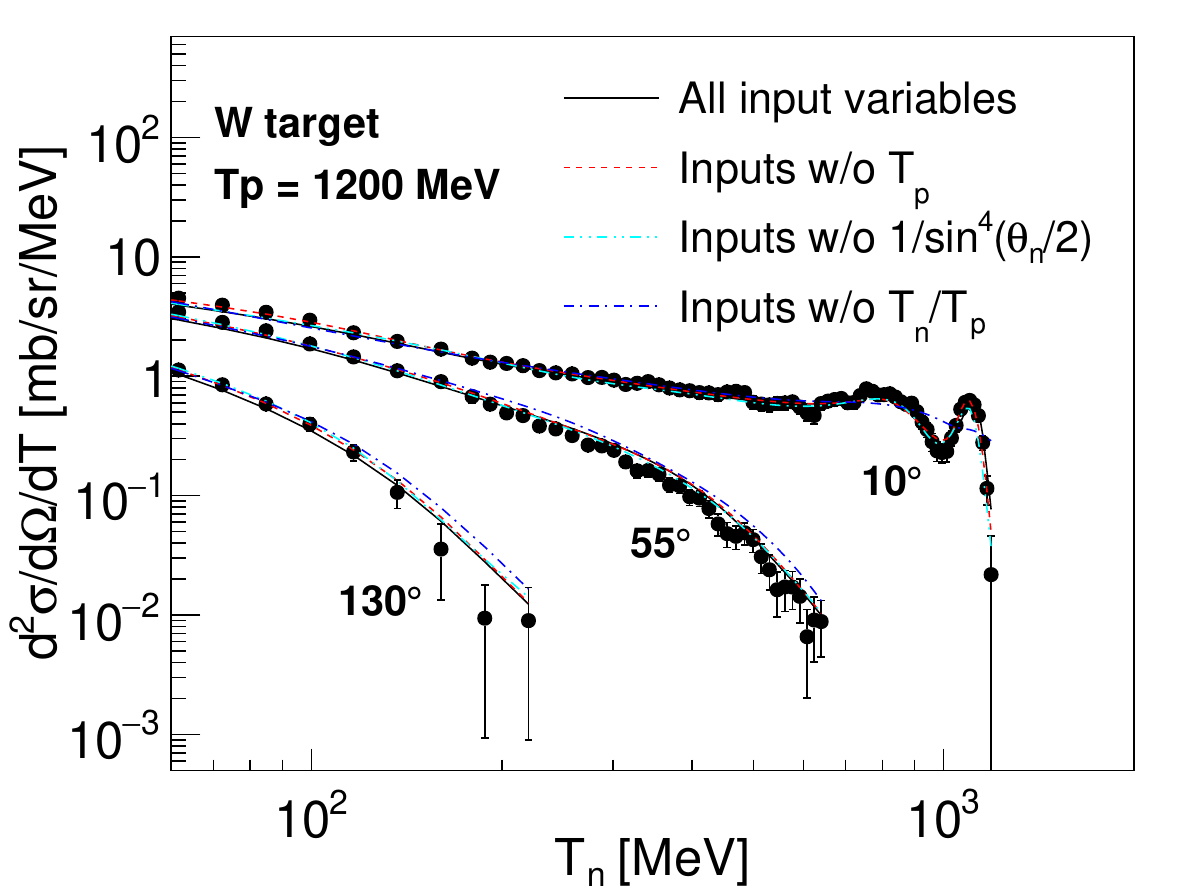}
\caption{
(Color online) ANN predictions for the neutron DDCS as a function of the spallation neutron energy using different input variables for the ANN framework: all input variables $\{\theta_n, 1/\sin^4(\theta_n), \log(T_n/T_p), (T_n/T_p), \log(T_p), T_p, A^{2/3}, N/Z\}$ (black solid curve), with only the variable $T_p$ removed (red dashed curve), with only the variable $1/\sin^4(\theta_n)$ removed (cyan dash-dot-dot curve), and with only the variable $T_n/T_p$ removed (blue dash-dotted curve). The ANN predictions are for the proton-induced spallation reaction on a W target at 1200 MeV, and they are compared with the experimental data \cite{Leray:2001pp}.
}
\label{fig:W_1200MeV_diffAngles_diffInputVars}
\end{center}
\end{figure}

We proposed eight input variables for the ANN framework of neutron DDCS modeling in Sec. \ref{sec:ann-framework}. Five of them ($\theta_n$, $\log(T_n/T_p)$, $\log(T_p)$, $A^{2/3}$, and $N/Z$) are indispensable because they are the minimum variables that define the conditions of the reaction channel. Logarithms of the proton and neutron energies are used because these energies cover several orders of magnitude. The additional input variable $1/\sin^4(\theta_n)$ is suggested to describe the angular dependence of the quasi-elastic peak in the neutron energy spectrum. The additional energy variables $(T_n/T_p)$ and $T_p$ are proposed to emphasize the weighting at high neutron and proton energies, respectively. This is important when the DDCS exhibits quite different behaviors in the low- and high-energy regions. Whether these three additional variables are necessary or redundant for a precise ANN model of DDCS is tested through comparative studies.

Fig. \ref{fig:Validation_losses_with_diff_input_variables} shows the validation losses of ANN models with different sets of input variables. The validation loss obtained using all eight input variables is taken as a reference. Removing either $T_p$ or $1/\sin^4(\theta_n)$ from the ANN input layer leaves the validation loss almost unchanged. This implies that $T_p$ and $1/\sin^4(\theta_n)$ are redundant for describing the neutron DDCS because $\log(T_p)$ and $\theta_n$ are already sufficient to govern the projectile-energy and neutron-angle dependencies of the DDCS. However, if $(T_n/T_p)$ is removed from the ANN input, the resulting validation loss is notably larger and stops decreasing after only 100 epochs. This means that $(T_n/T_p)$ is critical for precisely describing the neutron DDCS. Compared with $\log(T_n/T_p)$, the input variable $(T_n/T_p)$ gives much larger weight to neutrons in the high-energy tail. In nuclear spallation reactions, the neutron energy spectrum is very complicated, ranging from the low-energy evaporation region to the high-energy region of intense collisions. The variable $\log(T_n/T_p)$ emphasizes the description of the energy spectrum in the low-energy region, whereas $(T_n/T_p)$ emphasizes the description in the high-energy region. Both $\log(T_n/T_p)$ and $(T_n/T_p)$ are needed for a fine description of the complex neutron energy spectrum.

The importance of the input variable $(T_n/T_p)$ is also reflected in the reproduced neutron energy spectra. Fig. \ref{fig:W_1200MeV_diffAngles_diffInputVars} demonstrates the reproduced neutron DDCS as a function of neutron energy for ANN models with different combinations of input variables, compared with the tungsten data at 1200 MeV. If the variable $(T_n/T_p)$ is simply removed, the ANN cannot reproduce the quasi-elastic and quasi-inelastic peaks in the high-energy tail, yielding a much larger deviation from the data than the ANN with $(T_n/T_p)$ as an input variable. The quasi-elastic peak is located in and dominates the high-energy region ($T_n/T_p\lesssim 1$), where $(T_n/T_p)$ gives a large weight. This is why $(T_n/T_p)$ is vital for describing the quasi-elastic peak in the neutron energy spectrum well, in addition to $\log(T_n/T_p)$. The ANN prediction without either $T_p$ or $1/\sin^4(\theta_n)$ in the input layer still describes the neutron energy spectrum well. $T_p$ is redundant with respect to $\log(T_p)$, and $1/\sin^4(\theta_n)$ is redundant with respect to $\theta_n$. These conclusions regarding the neutron DDCS are consistent with the above discussion of the validation losses.

For the final ANN model, we discard the redundant variables $T_p$ and $1/\sin^4(\theta_n)$; the following six variables are used as the key input information for the ANN: $\theta_n$, $\log(T_n/T_p)$, $(T_n/T_p)$, $\log(T_p)$, $A^{2/3}$, and $N/Z$. How well the final ANN model agrees with the experimental measurements is presented in the following subsections.

\subsection{Comparisons between ANN model and training data}
\label{subsec:trained-data-comparision}

\begin{figure*}[htbp]
\begin{center}
\includegraphics[width=0.75\textwidth]{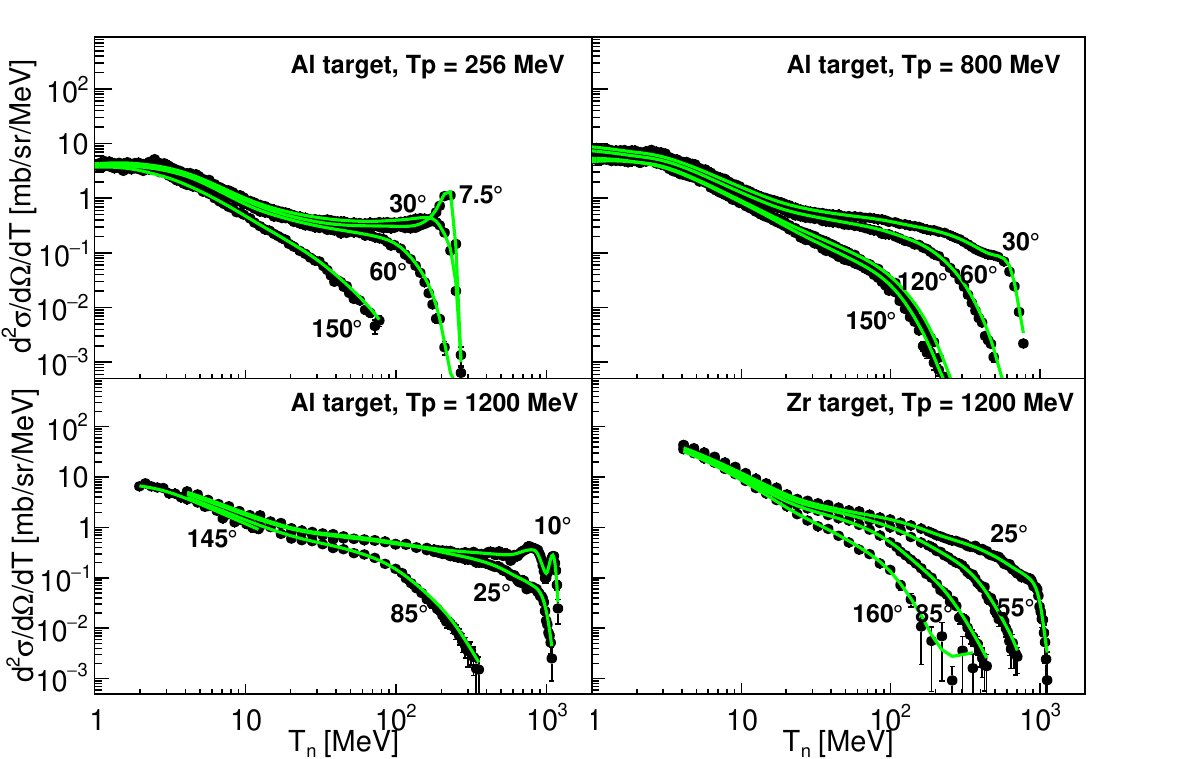}
\caption{
(Color online) Comparisons between the ANN model predictions and the experimental data for Al and Zr targets used in training \cite{Meier:1992anx,Amian:1992jal,Leray:2001pp}. The neutron emission angles are indicated in the plot.
}
\label{fig:DataComparision_Al_Zr}
\end{center}
\end{figure*}

\begin{figure*}[htbp]
\begin{center}
\includegraphics[width=0.75\textwidth]{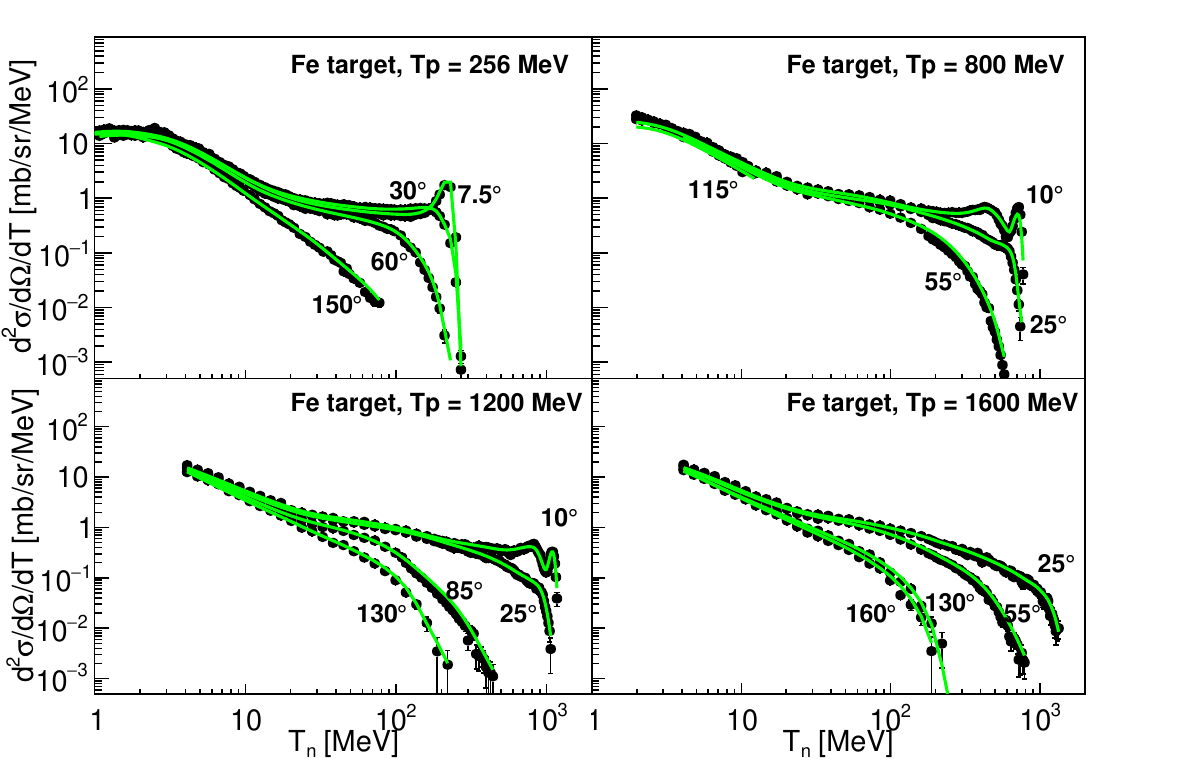}
\caption{
(Color online) Comparisons between the ANN model predictions and the experimental data for an Fe target used in training \cite{Meier:1992anx,Amian:1992jal,Leray:2001pp}. The neutron emission angles are indicated in the plot.
}
\label{fig:DataComparision_Fe}
\end{center}
\end{figure*}

\begin{figure*}[htbp]
\begin{center}
\includegraphics[width=0.75\textwidth]{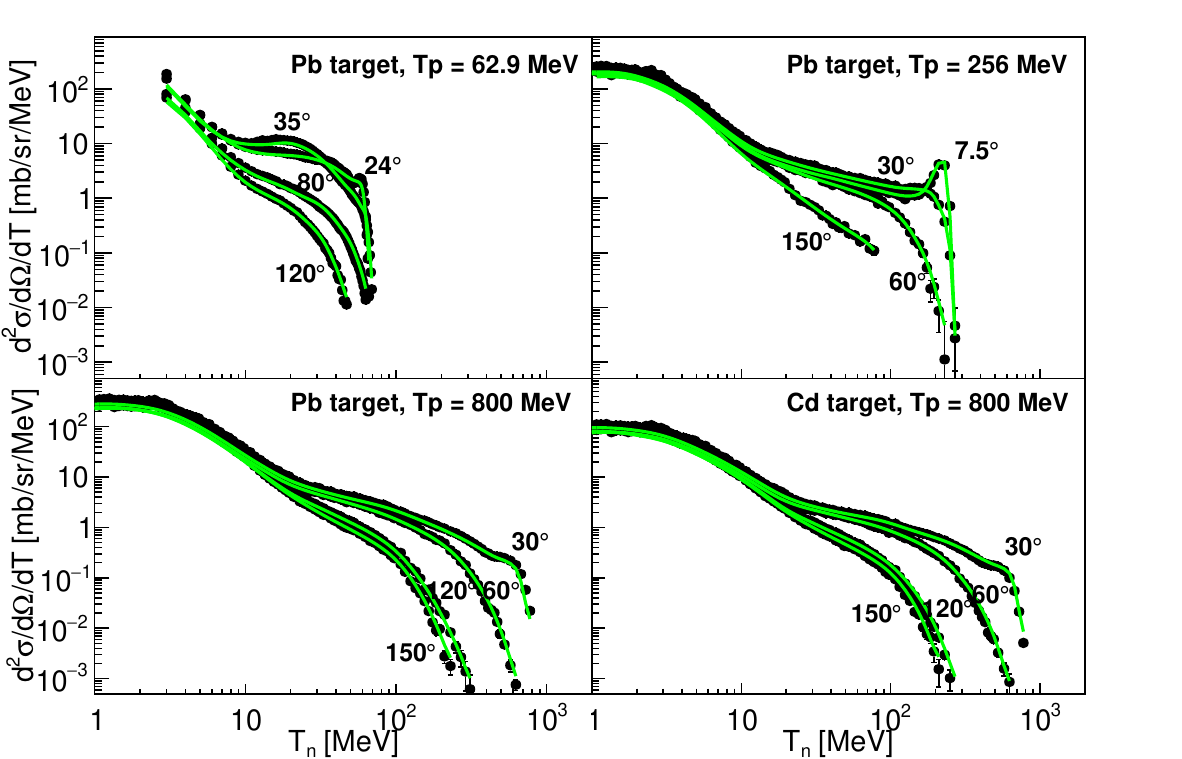}
\caption{
(Color online) Comparisons between the ANN model predictions and the experimental data for Pb and Cd targets used in training \cite{Meier:1992anx,Amian:1992jal,Leray:2001pp,Guertin:2004}. The neutron emission angles are indicated in the plot.
}
\label{fig:DataComparision_Pb_Cd}
\end{center}
\end{figure*}

\begin{figure*}[htbp]
\begin{center}
\includegraphics[width=0.75\textwidth]{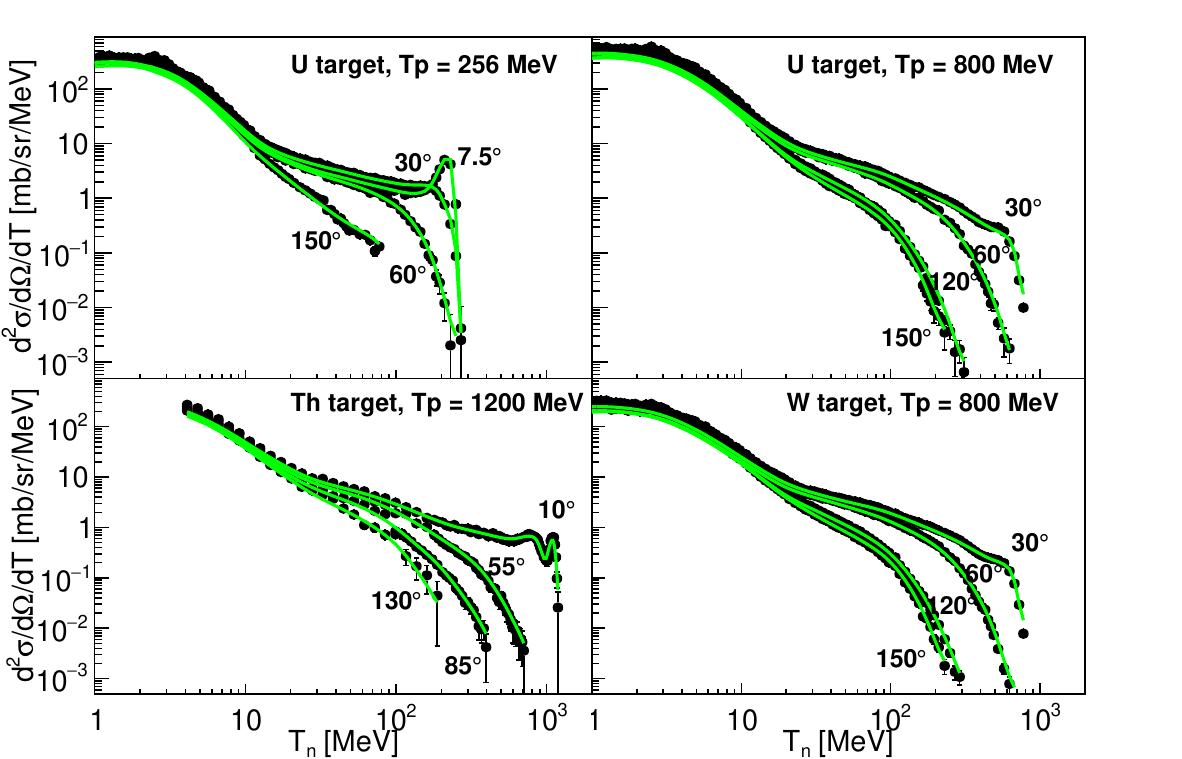}
\caption{
(Color online) Comparisons between the ANN model predictions and the experimental data for U, Th, and W targets used in training \cite{Meier:1992anx,Amian:1992jal,Leray:2001pp}. The neutron emission angles are indicated in the plot.
}
\label{fig:DataComparision_U_Th_W}
\end{center}
\end{figure*} 

For the final ANN model, only six key input variables are used, and no regularization is applied during training. ANN training was stopped at 300 epochs, where the validation loss is well stabilized. The trained ANN model parameters are saved for predicting neutron DDCS for any nuclear target under any kinematical conditions.

For a closure test, the DDCS predictions of the ANN model are compared with the experimental data used for training the ANN. Fig. \ref{fig:DataComparision_Al_Zr} shows comparisons between the ANN model predictions and the training data for aluminum and zirconium targets. Fig. \ref{fig:DataComparision_Fe} shows comparisons for an iron target. Fig. \ref{fig:DataComparision_Pb_Cd} shows comparisons for lead and cadmium targets. Fig. \ref{fig:DataComparision_U_Th_W} shows comparisons for uranium, thorium, and tungsten targets. One sees that the obtained small-scale ANN model reproduces the experimental data excellently. Note that the experimental data were taken from several different groups measured at different facilities worldwide.

It is observed in Figs. \ref{fig:DataComparision_Al_Zr} - \ref{fig:DataComparision_U_Th_W} that the DDCS data for various nuclei from Al to U are all well reproduced. We conclude that the nuclear dependence of the neutron DDCS is well described by the trained ANN model. The neutron DDCS at different neutron angles are presented in the figures, with a minimum angle of $7.5^\circ$ and a maximum angle of $160^\circ$. The high-energy neutron distributions decrease quickly as the neutron angle increases. All patterns of the angular dependence of the neutron DDCS are satisfactorily reproduced by the ANN model. It is also found that the ANN model successfully describes the DDCS dependence on projectile energy. The data for Al, Fe, and Pb were measured at several different incident energies. In Fig. \ref{fig:DataComparision_Fe}, one clearly sees that the maximum neutron energy increases with incident energy. Meanwhile, the DDCS value at a certain angle increases slowly with increasing projectile energy. The dependencies of the neutron DDCS on nuclear target, incident energy, and neutron emission angle are all excellently captured by the ANN model.

The patterns of the neutron energy spectrum are complicated, with a gentle slope in the low-energy region, power-law behavior in the intermediate-energy region, and a fast drop near the high-energy end of the spectrum. The low-energy neutron distribution of evaporation neutrons has almost no angular dependence, whereas the high-energy distribution of prompt neutrons has a strong angular dependence. In general, all patterns of the neutron energy spectrum are reproduced by the ANN model mostly within the experimental uncertainties. It is worth mentioning that the quasi-elastic peak ($p+A\rightarrow p+n+X$) and the quasi-inelastic peak ($p+A\rightarrow \Delta+n+X$, at high incident energy) are well described by the proposed ANN model. For example, the DDCS for the Fe target at $7.5^\circ$ with a 256 MeV proton beam displays a quasi-elastic peak, and the DDCS for the Fe target exhibits quasi-elastic and quasi-inelastic peaks at $10^\circ$ with high-energy proton beams (800 and 1200 MeV).

One may find in Figs. \ref{fig:DataComparision_Pb_Cd} and \ref{fig:DataComparision_U_Th_W} that the ANN model predictions slightly underestimate the neutron DDCS in the low-neutron-energy region for tungsten, lead, and uranium targets. Note that the experimental neutron DDCS data for tungsten, lead, and uranium targets span more than five orders of magnitude at different neutron energies. It is very difficult to precisely describe a quantity that changes over more than five orders of magnitude using a simple model. The minor underestimation of the neutron DDCS in the low-energy region around 1 MeV for the tungsten, lead, and uranium targets may be due to the limitations of the small-scale ANN model used. This tiny deviation does not negate the fact that the ANN model successfully describes the DDCS as functions of multiple variables over wide ranges.

\subsection{Comparisons between ANN model and unseen data}
\label{subsec:untrained-data-comparision}

\begin{figure*}[htbp]
\begin{center}
\includegraphics[width=0.75\textwidth]{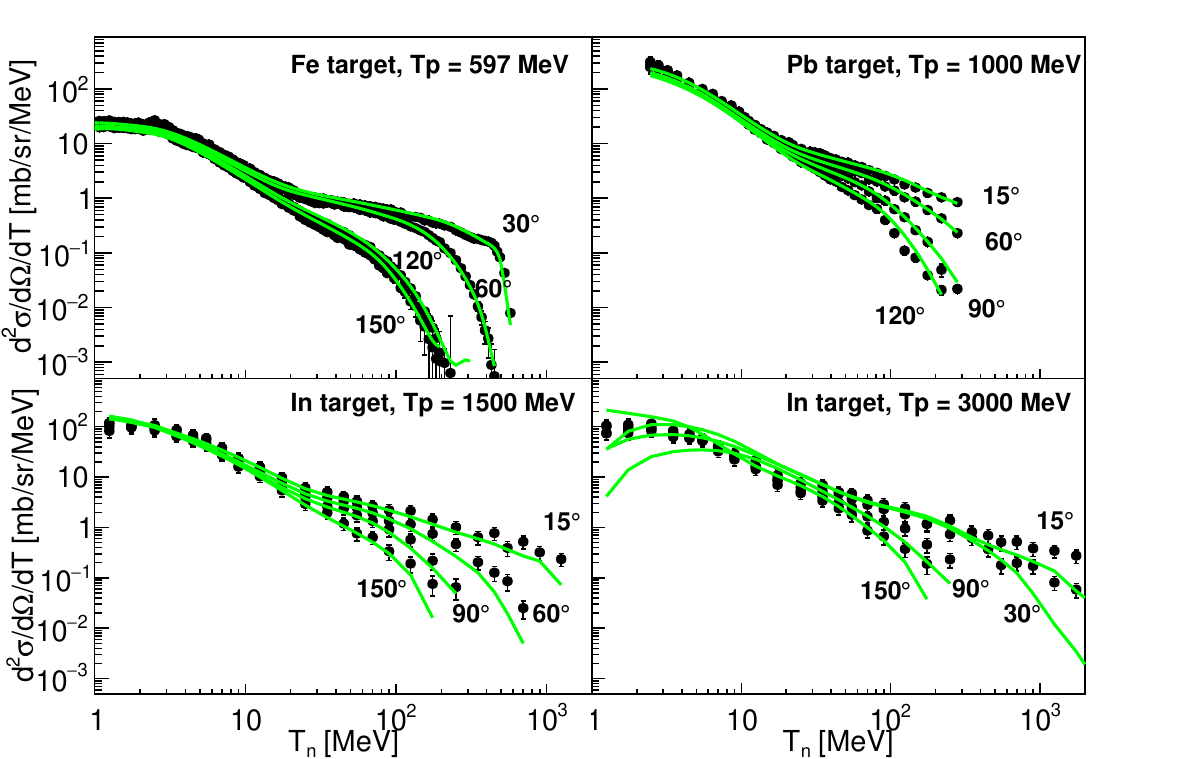}
\caption{
(Color online) Comparisons between the ANN model predictions and the experimental data for Fe, Pb, and In targets not used in the ANN training (unseen data for the ANN) \cite{Amian:1993dhv,Trebukhovsky:2003uh,Ishibashi:1997gbe}. The neutron emission angles are indicated in the plot.
}
\label{fig:Predictions_for_untrained_data}
\end{center}
\end{figure*} 

Further testing with unseen data is conducted for the trained ANN model. In Fig. \ref{fig:Predictions_for_untrained_data}, the ANN model predictions are compared with experimental data outside the training data set (unseen data for the model). This type of validation is a direct and rigorous test of the generalization ability of the ANN model.

The incident energies 597, 1000, 1500, and 3000 MeV of the test data in Fig. \ref{fig:Predictions_for_untrained_data} are not included in the training data. For example, the closest energies to 597 MeV in the training data are 256 and 800 MeV. The good agreement between the model predictions and the unseen data at 597, 1000, and 1500 MeV indicates that the proposed ANN model has strong extrapolation power for predicting neutron DDCS at different incident energies. The extrapolation power is restricted to an incident-energy region not far from the range covered by the training data. The extrapolation of the DDCS to an incident energy of 3000 MeV does not agree well with the unseen data, especially in the low-energy region ($T_n\lesssim 5$ MeV) and at the high-energy tail. 3000 MeV is far from the maximum projectile energy included in the training data.

The extrapolation power of the ANN model to an arbitrary neutron angle is also tested with unseen data. Note that there are no DDCS data at a neutron angle of $90^\circ$ in the training data. Hence, the neutron energy spectra at $\theta_n=90^\circ$ in Fig. \ref{fig:Predictions_for_untrained_data} are extrapolated by the ANN model from training data at different angles and different projectile energies. One finds that the extrapolation ability of the ANN model is also powerful in providing predictions at a different neutron angle.

From comparisons of the ANN model predictions with unseen data, one can also verify the extrapolation power of the model to different nuclear targets. There are no data for an indium target in the training data. Thus, the predictions for the In target in Fig. \ref{fig:Predictions_for_untrained_data} are completely extrapolated in the space of $(A^{2/3}, N/Z)$ by the obtained ANN model. The DDCS predictions for the In target at an incident energy of 1500 MeV are consistent with the unseen experimental data mostly within the experimental errors. This is a quite important capability of a model for making complete descriptions of nuclear reactions for all nuclides. The extrapolation to 3000 MeV is not accurate for the reasons discussed above.

The proposed small-scale ANN model has good generalization ability, as judged from the extrapolation results discussed above. It is accurate and precise for predicting nuclear reaction DDCS in a domain not far from the region covered by the training data. Note that some of the test data were measured by groups different from those providing the training data. Therefore, the agreement between the ANN model predictions and the unseen data is not only evidence of the generalization ability of the ANN but also an indication of consistency among different experimental measurements.

\subsection{ANN model predictions}
\label{subsec:ann-prediction-demo}

\begin{figure*}[htbp]
\begin{center}
\includegraphics[width=0.75\textwidth]{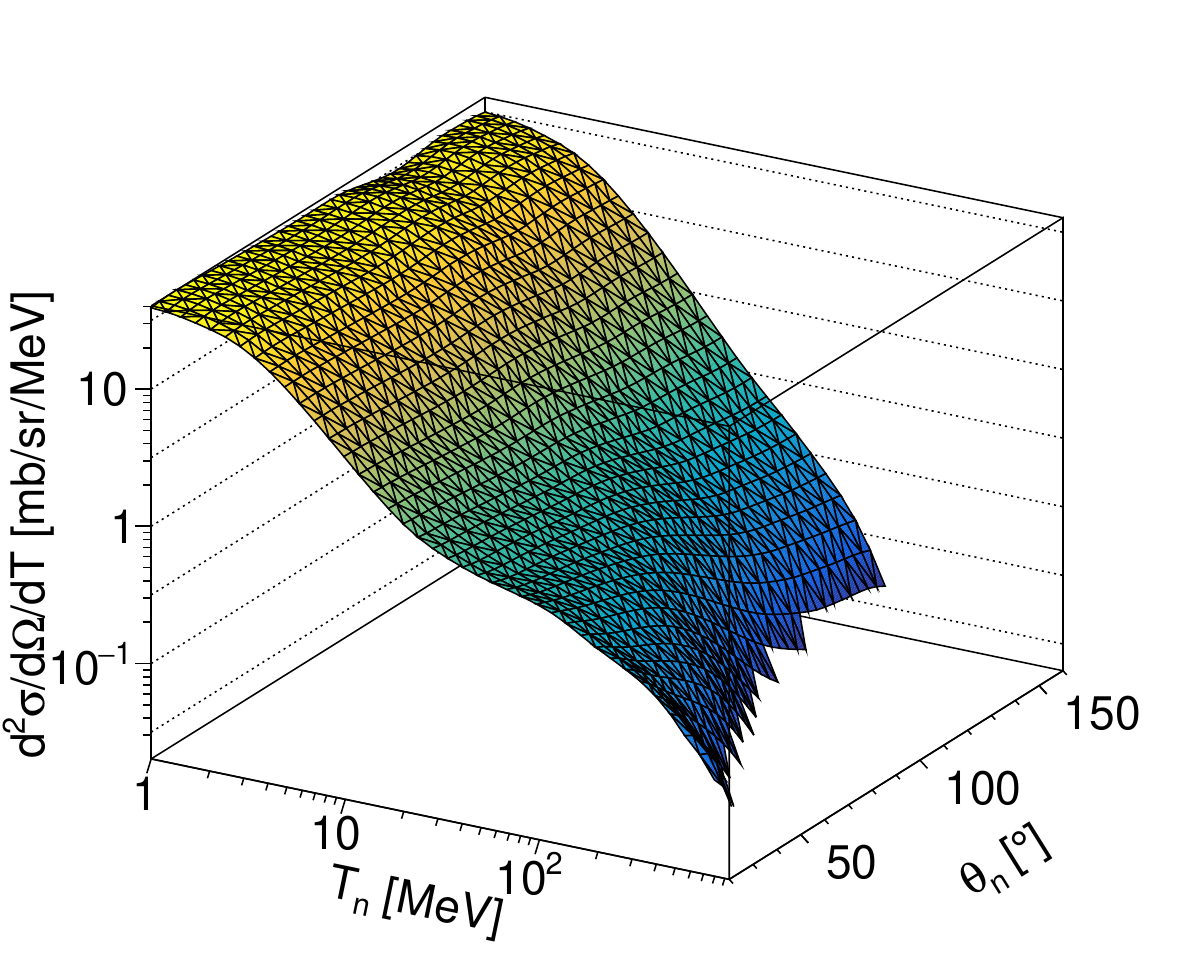}
\caption{
(Color online) ANN model prediction of the neutron DDCS for a Cu target as functions of the neutron energy $T_n$ and neutron emission angle $\theta_n$. The incident proton energy is 1000 MeV.
}
\label{fig:PredictionCu_DDCS}
\end{center}
\end{figure*}

\begin{figure*}[htbp]
\begin{center}
\includegraphics[width=0.75\textwidth]{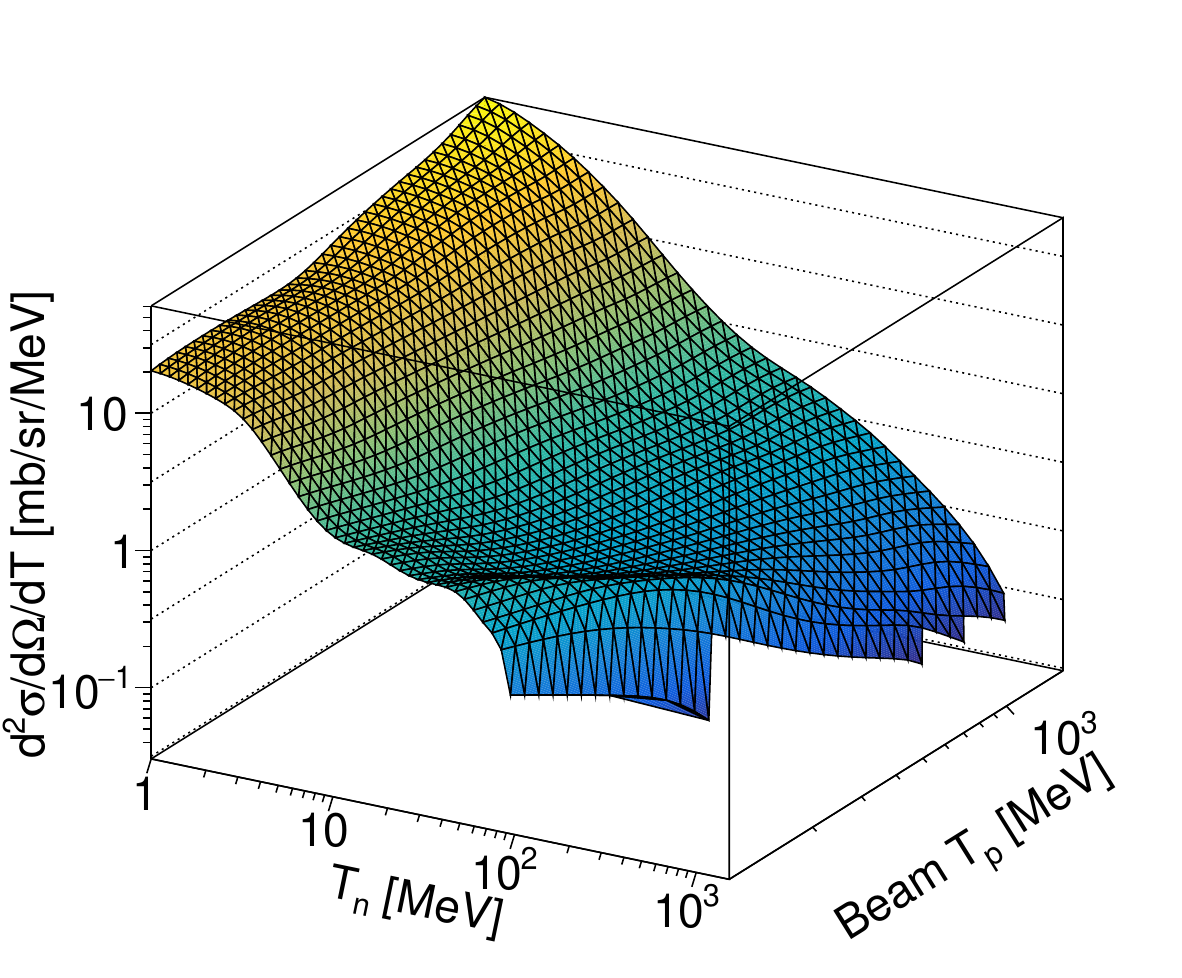}
\caption{
(Color online) ANN model prediction of the neutron DDCS for a Cu target as functions of the emitted neutron energy $T_n$ and incident proton energy $T_p$. The neutron emission angle is fixed at $30^\circ$.
}
\label{fig:PredictionCu_DDCS_vs_Tp}
\end{center}
\end{figure*} 

Detailed DDCS predictions for a copper target have been made for further future tests of the ANN model. The experimental data for the Cu target are not included in the training data. As a result, testing with neutron DDCS data for the Cu target is significant for further testing the generalization capability of the model.

Fig. \ref{fig:PredictionCu_DDCS} shows the neutron DDCS of spallation reactions on a copper target as functions of neutron energy $T_n$ and angle $\theta_n$. One sees a fast decrease in the DDCS with increasing neutron energy, and the neutron energy spectrum varies smoothly with the emission angle $\theta_n$. The high-energy tail becomes shorter as the neutron angle increases. Fig. \ref{fig:PredictionCu_DDCS_vs_Tp} displays the neutron DDCS at $\theta_n=30^\circ$ as functions of projectile energy and emitted neutron energy. At relatively low projectile energy, the quasi-elastic peak at $\theta_n=30^\circ$ is more apparent. In general, the overall neutron DDCS increases with increasing incident energy, especially for the DDCS in the low-energy region $T_n\lesssim 10$ MeV. The neutron DDCS given by the ANN model is a continuous and smooth function in the multidimensional kinematic space of $\theta_n$, $T_n$, and $T_p$, with no unphysical or anomalous behavior found.

\section{Summary and outlook}\label{sec:summary}

A model for neutron emission in nuclear spallation reactions is a critical and demanding tool for many applications, such as ADANES for nuclear waste transmutation and nuclear fuel generation, spallation neutron sources, rare-isotope production, radiation shielding analysis, and medical and astrophysical applications. Some potentially missing mechanisms need to be implemented to improve the traditional intranuclear cascade or quantum molecular dynamics models. Tuning a large set of parameters to reproduce all experimental observables is very challenging for dynamical models. Therefore, it is very difficult to construct a high-precision model over a wide kinematic region by simply updating dynamical reaction models. Instead of improving dynamical reaction models, we have explored a data-driven model based on machine learning. Because the data-driven nuclear reaction model relies solely on experimental data, the machine-learning method can precisely reproduce the experimental data. The ANN framework is inherently unbiased and highly flexible in modeling complicated multidimensional functions, and it has proven very successful in reproducing and predicting the neutron DDCS for various nuclear targets over a broad kinematic region.

In this study, we find that the proposed small-scale ANN ($128\times128\times128$) reproduces the experimental data used in training well within the experimental uncertainties, owing to the high flexibility of the ANN. The projectile-energy, neutron-energy, and angular dependencies of the DDCS are all well captured by the trained ANN model. The trained ANN model is also found to have satisfactory generalization ability. It is feasible to construct a high-precision data-driven model of nuclear spallation using the machine-learning technique of the ANN framework. Using uncertainty-incorporated replicas of the experimental data helps prevent overfitting during ANN training, and a large number of data replicas are recommended. With the commonly used L2 and dropout regularization schemes, the sharp oscillations of the DDCS in the quasi-elastic and quasi-inelastic peaks cannot be reproduced. Either the L2 or the dropout regularization scheme yields a rather flat functional form in the quasi-elastic and quasi-inelastic regions. From the ANN machine-learning analysis, we found that the six variables $\theta_n$, $\log(T_n/T_p)$, $(T_n/T_p)$, $\log(T_p)$, $A^{2/3}$, and $N/Z$ are the key input variables for the ANN framework. Since the neutron energy spectrum is a complicated function of neutron energy, both $\log(T_n/T_p)$ and $(T_n/T_p)$ are necessary as ANN inputs to achieve a precise description of all details of the spectrum.

Predictions from the final ANN model are provided for further tests. Owing to the limited scope of the training data used in this work, the obtained ANN model is recommended for predicting neutron DDCS in the following kinematic region for heavy nuclei: projectile energies from 100 to 1600 MeV, neutron angles from 0 to 180 degrees, emitted neutron energies from 0 to the incident energy, and target nuclei from $^{27}$Al to $^{238}$U. Predictions outside this suggested range can still be made by the ANN model; however, the precision of the results is not guaranteed. 

The ANN model for spallation reaction cross sections, trained on sufficient data, has numerous potential applications in nuclear energy science and engineering, including assessing the consistency of DDCS data among different measurements in similar kinematic regions, benchmarking nuclear reaction models, and building a high-energy nuclear data library. A more direct application is the implementation of the obtained ANN model in particle transport codes to reweight Monte-Carlo sampling events. It is anticipated that reweighting simulated events with the high-precision ANN model will reduce model biases and uncertainties within the applicable kinematic domain. Further studies are warranted to explore the benefits of applying a properly trained ANN model in nuclear computation, simulation and engineering. 

The data-driven ANN model relies heavily on experimental data. We look forward to more high-precision experimental measurements to improve the ANN model within the machine-learning framework. In addition to the DDCS for neutron production, the DDCS for other particle production are also quite valuable and are essential for building an ANN model that provides a complete description of spallation reactions. In the future, a large-scale ANN model can be explored using much more experimental data.

\begin{acknowledgments}
This work is partly supported by the National Development and Reform Commission of China (Large Research Infrastructures of 12th Five-Year Plan: China initiative Accelerator Driven System, No. 2017-000052-75-01-000590)
\end{acknowledgments}

\bibliographystyle{apsrev4-1}
\bibliography{list-of-refs}

\end{document}